\documentclass[aps,pre,showpacs,eqsecnum]{revtex4}
\usepackage{amsmath,bm}
\usepackage{graphicx,color}

\begin{document}
\sloppy \large

\title{Vorticity statistics in the direct cascade of two-dimensional turbulence}

\author{Gregory Falkovich$^{1}$, Vladimir Lebedev$^{2}$ and Mikhail Stepanov$^{3}$}

\date{\today}

 \begin{abstract}

For the steady-state direct cascade of two-dimensional ($2d$) Navier-Stokes turbulence,
we derive analytically the probability of strong vorticity fluctuations. When $\varpi$ is
the vorticity coarse-grained over a scale $R$, the probability density function (PDF),
${\cal P}(\varpi)$, has an asymptotic behavior $\ln{\cal P}\sim - \varpi/\varpi_{rms}$ at
$\varpi\gg\varpi_{rms}= [H\ln(L/R)]^{1/3}$, where $H$ is the enstrophy flux and $L$ is
the pumping length. Therefore, the pdf has exponential tails and it is self-similar, i.e.
it can be presented as a function of a single argument, $\varpi/\varpi_{rms}$, in
distinction from other known direct cascades.

 \end{abstract}


 \affiliation{
 $^1$ Physics of Complex Systems, Weizmann Institute of Science, Rehovot 76100 Israel \\
 $^2$ Landau Institute for theoretical physics RAS, 119334, Kosygina 2, Moscow, Russia  \\
 $^3$ Dept. of Mathematics, University of Arizona, Tucson, AZ 85721 USA}

 \maketitle

\section{Introduction}

Turbulence is a paradigmatic far-from-equilibrium state of matter and the central
question of physics of turbulence is that of universality: how much one needs to know
about an external forcing (or initial data for decaying turbulence) to predict the basic
features of flow statistics. A related question is that of symmetries of the statistics,
particularly whether scale invariance appears for the scales distant from $L$, where
turbulence is excited \cite{Frisch,FS}. One distinguishes direct and inverse cascades
occurring at the scales much smaller or much larger than $L$, respectively. Data suggest
that the statistics of inverse cascades are scale invariant \cite{FS,FGV,F}. For example,
the probability density function (PDF) ${\cal P}$ of $\varpi$ that is the vorticity
$\omega$ coarse-grained over a scale $R$ is empirically found to be a function of a
single variable rather than two in $2d$ inverse energy cascade: ${\cal P}(\varpi)
=\varpi^{-1} f(\varpi R^{-2/3})$ \cite{Tab,KG,EREC,BCV}. Such self-similarity was never
observed in direct cascades for whatever small $R/L$, the probability distributions of
$\varpi$ change their forms with varying the ratio $R/L$ \cite{Frisch,FS,FGV,F}.

One way to explain this profound difference is to argue that fluid motions are slower
when the scales are larger. As an inverse cascade proceeds, it has an ample time to be
effectively averaged over the small-scale fluctuations including those of the pumping,
whose only memory left is the value of flux it generates. On the contrary, small-scale
fast fluctuations in a direct cascade stay sensitive to the statistics of fluctuations at
larger scales \cite{LL}, nonlinearity then enhances the effect of fluctuations down the
cascade so that the small-scale statistics is dominated by rare strong fluctuations.

One can also explain the difference between the direct and inverse cascades
using the Lagrangian language. Correlation functions are accumulated along
Lagrangian trajectories (for the forced turbulence) or originate from initial
data transported along the Lagrangian trajectories (for the decaying
turbulence). Correlation functions are then proportional to the time the
trajectories spend within the volume of size $L$. When the trajectories
approach each other back in time, correlations appear at larger and larger
scales, which corresponds to an inverse cascade. In this case, two-particle
behavior effectively determines evolution of multi-particle configurations and
the second moment exponent determines the scaling of higher moments. Just the
opposite, direct cascades correspond to trajectories separating back in time,
one then relates the breakdown of scale-invariance at vanishing viscosity to
non-uniqueness of explosively separating trajectories in a non-smooth velocity
field; exponents of higher moments are then related to the law of decay of the
fluctuations of the shapes of multi-particle configurations which depend on the
number of particles \cite{FGV,CFG}. Say, for the passive scalar, the exponents
are independent of the scalar pumping but are dependent on the mixing velocity
statistics \cite{CFL}. Still, one needs to know an infinite number of
forcing-related parameters to predict the scalar statistics at however small
scales. For the direct cascades in Navier-Stokes turbulence and similar
nonlinear problems, it is not even known whether the exponents of the velocity
moments are universal or not.

Prior knowledge was based on experimental and numerical data, the only analytical results
were obtained for passive fields in synthetic flows \cite{CFKL,SS,GK,FGV,BL}. Here, for
the first time, the vorticity PDF tail is analytically derived from the equation of
motion.

We consider the direct (enstrophy) cascade of $2d$ turbulence \cite{Kra67,Leith,Batch},
which in Lagrangian terms is peculiar since it corresponds to an exponential separation
of trajectories. Indeed, the physical mechanism of the cascade is that pumping-produced
vorticity blobs are deformed by the flow into thin streaks: stretched in one direction
and contracted in another one until viscosity dissipates them, like for a passive scalar.
An important distinction from the passive scalar is that the vorticity $\omega$ is
related to the flow velocity $\bm v$: $\omega=\mathrm{curl}\, \bm v$. Constancy of the
enstrophy flux over scales, $\langle({\bm v}_1\cdot\nabla_1+{\bm v}_2\cdot\nabla_2)
\omega_1\omega_2\rangle=\mathrm{const}$, suggests the scaling $|\bm v(\bm r)-\bm
v(0)|\propto r$ i.e. spatially smooth velocity. In a steady state, the enstrophy
dissipation $\nu|\nabla\omega|^2$ must stay finite in the inviscid limit $\nu\to 0$. The
velocity then cannot be perfectly smooth, but the possible singularities are no stronger
than logarithmic \cite{FL,Eye}. If one assumes self-similarity in a sense that the PDF of
the coarse-grained vorticity is ${\cal P}(\varpi) =\varpi^{-1} f[\varpi^a /\ln(L/R)]$,
then the flux constancy requires $a=3$ \cite{Kra67,FL,Claudia}.

There are some consequences of the self-similarity. Say, the enstrophy transfer
time through a given scale $R$, determined by the stretching/contraction rate,
can be estimated as a turn-over time or an inverse vorticity at this scale. On
the one hand, that time decreases with the scale as $\ln^{-1/3}(L/R)$, which
would suggest that the small-scale statistics is sensitive to the statistics at
larger scales. On the other hand, the total time of enstrophy transfer from $L$
down to the viscous scale $\eta$ diverges $\propto\ln^{2/3}(L/\eta)$ as
$\eta\to0$. Particle trajectories are then expected to separate exponentially
rather than explosively and stay unique even in the inviscid limit, that makes
self-similarity plausible, according to the above Lagrangian arguments. Note
that von Neumann \cite{Neu} and  Kraichnan \cite{Kra67} argued that an infinite
number of vorticity conservation laws can make the vorticity cascade
non-universal, yet Falkovich and Lebedev later argued that the fluxes of higher
vorticity invariants must be irrelevant due to the phenomenon of ``distributed
pumping" \cite{FL}. Recently, self-similarity breakdown was found empirically
for the vorticity isolines, which are conformal invariant in the inverse
cascade, while in the direct cascade they are not scale-invariant but
multi-fractal with the fractal dimension $3/2$ and higher dimensions saturating
at $1$ \cite{BBCF,BBCF07} (that may be related to strain persistence that leads
to vorticity organized in long thin streaks). That makes it natural to expect
that the bulk vorticity statistics is not self-similar as well.

The present work is devoted to analytical description of a single-time vorticity
statistics in the steady-state $2d$ turbulence in the direct cascade. We analytically
derive the non-Gaussian tail of the PDF of $\varpi$, that is the vorticity coarse-grained
over the scale $R$, in the direct (enstrophy) cascade. We show that the tail is
exponential,
 \begin{eqnarray}
 \ln{\cal P}(\varpi) \sim
 -\frac{|\varpi|}{[H\ln(L/R)]^{1/3}}  \,,
 \label{answer0}
 \end{eqnarray}
for a driving force with a finite correlation time. In particular, Eq. (\ref{answer0})
shows that the vorticity PDF is self-similar, i.e. it can be presented as ${\cal
P}(\varpi)= \varpi^{-1} f[\varpi^{3} /\ln(L/R)]$. Moreover, up to the order-unity factor,
the tail (\ref{answer0}) is determined by a single parameter, $H$, that is the flux of
the squared vorticity, which also determines the $\varpi_{rms}$ i.e. the bulk of the pdf.
To obtain the single-point vorticity PDF, the ratio $L/R$ should be substituted by
$\sqrt{Re}$ in Eq. (\ref{answer0}).

The structure of our paper is as follows. .......................

 \section{Basic equations}
 \label{sec:basic}

The incompressible $2d$ Euler equation can be written for the vorticity
$\omega=\partial_x v_y-\partial_y v_x$:
 \begin{equation}
 \frac{\partial\omega}{\partial t}
 + ({\bm v}\nabla) \, \omega = \phi \,.
 \label{Eqw}
 \end{equation}
Here $\phi$ is curl of the external force $\bm f$ exciting the turbulence: $\phi=
\partial_x f_y -\partial_y f_x$. The viscous term is omitted in (\ref{Eqw}), which means
that we consider flow variations on scales much larger than the viscous scale $\eta$.
We shall describe the flow in the Lagrangian reference frame attached to a fluid particle
placed at the origin, such that ${\bm v}({\bm 0}) = {\bm 0}$. Then the velocity is
expressed via the vorticity as
 \begin{equation}
 v_\alpha({\bm r}) = -\epsilon_{\alpha\beta}
  \int \frac{d^2 r'}{2\pi}
  \left( \frac{r_\beta - r'_\beta}{|{\bm r}-{\bm r'}|^2} +
  \frac{r'_\beta}{|\bm r'|^2} \right) \omega(\bm r') \,.
 \label{Vel}
 \end{equation}
The pumping $\phi$ is assumed to be a random Gaussian field spatially correlated on the
scale $L$ and short correlated in time. Then its variance is $\langle \phi(0,{\bm 0})
\phi(t,{\bm r})\rangle =\delta(t)\chi(r)$, where $\chi(r)$ rapidly tends to zero as $r$
exceeds $L$. As we shall see below, the processes that contribute to the vorticity PDF
tails take a long time which allows effective averaging over forcing so that our results
are asymptotically valid for any forcing with a finite correlation time.

The statistics of the flow can be examined within the framework of Martin-Siggia-Rose
formalism \cite{MSR,D,J,DP} so that all the averages (correlation functions)
characterizing the flow are calculated as functional integrals, $\int {\cal D}p\, {\cal
D}\omega\, \exp(i{\cal I})\dots$, with the effective action
 \begin{eqnarray}
 {\cal I}=\int dt\, d^2r\,
 p({\bm r})\left[\partial_t\omega\!
 + \!{\bm v}\nabla\omega \!+\!\frac{i}{2}\int\!  d^2r'\,
 \chi(|{\bm r}-{\bm r}'|)p({\bm r}')\right] \,.
 \label{actionc}
 \end{eqnarray}
Here $p$ is an auxiliary field introduced to put the equation of motion (\ref{Eqw}) into
the exponent. Since the action (\ref{actionc}) contains a cubic term originating from the
nonlinear term in Eq. (\ref{Eqw}), one is unable to calculate the functional integrals
explicitly. Nor one is able to treat the third-order term by a perturbation theory, since
there is no small parameter in the expansion. In other words, we deal with the theory
where the coupling is strong. What allows for an analytic description is that we consider
rare strong fluctuations i.e. describe tails of the vorticity PDF.

We consider the PDF of $\varpi$, that is the vorticity $\omega$ coarse-grained over a
scale $R$ from the interval of the direct cascade, that is we assume that $R$ is much
smaller than the pumping scale $L$ but larger than the viscous length $\eta$. A general
strategy to find tails of the PDF, ${\cal P}(\varpi)$, is to calculate the corresponding
functional integral in the saddle-point approximation utilizing  the ratio
$\varpi/\varpi_{rms}$ as a large parameter. The way to do that is the so-called instanton
formalism adapted for turbulence problems \cite{BL,FKLM,GM,BFKL,Chert}. In this way, one
looks for an extremum of the action (\ref{actionc}), defined by the instanton (extremum)
equations $\delta {\cal I}/\delta\omega=0=\delta {\cal I}/\delta p$ with appropriate
boundary conditions. Both the action and the measured quantity $\omega$ are invariant
with respect to rotations and so are instanton equations and their boundary conditions.
However, axial symmetry turns nonlinear terms in the instanton equations into zero
killing dynamics. In other words, a ``naive instanton'' is meaningless. The physical
reason is quite transparent: there is neither stretching nor contraction for axially
symmetric flows so that the force can pump the vorticity forever. That means that flow
realizations that determine a given large value of $\varpi$ must have their axial
symmetry broken. We establish below that the angle-dependent part of the vorticity
realizations remains much smaller than the isotropic part during most of the evolution
(by virtue of the large parameter $\varpi /\varpi_{rms}$). That will allow us to
integrate over angular degrees of freedom (in the Gaussian approximation) and obtain a
renormalized action for the zero harmonic $\omega_0$. Moreover, we show that only the
second angular harmonic provides for the relevant renormalization by virtue of the large
parameter $\ln(L/r)$. We then find the new (effectively axially symmetric) instanton that
corresponds to the renormalized action and gives the tail of the coarse-grained vorticity
PDF, ${\cal P}(\varpi)$.

 \section{Separation of harmonics}
 \label{sec:separ}

We use polar coordinates, $x=r\cos\varphi$, $y=r\sin\varphi$, and expand the fields
$\omega$ and $p$ over the angular harmonics:
 \begin{equation}
 \omega(t,{\bm r})=\sum \omega_m(t,r)\exp(im\varphi), \quad
 {2\pi} p(t,{\bm r}) =\sum p_m(t,r)\exp(im\varphi).
 \label{harmonics}
 \end{equation}
Then the effective action (\ref{actionc}) splits into a number of terms ${\cal I}={\cal
I}_0 +\sum_{m>0}({\cal I}_m+{\cal I}_{-m}+{\cal I}_{mm} +{\cal I}_{im})+{\cal I}_3 $,
where ${\cal I}_0$ contains only the zero harmonics $\omega_0,p_0$. The last term ${\cal
I}_3$ is a sum of the third order terms containing harmonics with $m\neq 0$, it  is
neglected in what follows, which is justified below. The terms quadratic in
$p_m,\omega_m$ are written as follows
 \begin{eqnarray}
 {\cal I}_0=\int dt\,dr\, r \, p_0\partial_t \omega_0
 +\frac{i}{2} \int dt\, dr\, r \, d r'\,r'\,
 \chi_0(r,r') p_{0}(r) p_0(r') \,,
 \label{actionzero} \\
 {\cal I}_m=\int dt\,dr\, r \, p_{-m}
 \left\{\partial_t\omega_m+ {v_0}im\omega_m/r
 +\partial_r\omega_0\, v_{rm}\right\} \,,
 \label{actionm} \\
 {\cal I}_{im}=-\int dt\, dr\, r \,
 \partial_rp_0 \left(v_{r,m}\omega_{-m}
 +v_{r,-m}\omega_{m}\right) \,,
 \label{actioni} \\
 {\cal I}_{mm}=i \int dt\, dr\, r \, d r'\,r'\,
 \chi_m(r,r') p_{-m}(r) p_m(r') \,,
 \label{actionmm} \\
 \chi_m(r,r')=\int\!\frac{d\varphi}{2\pi}\,
 e^{im\varphi}\chi\left(\sqrt{r^2+r'^2
 -2rr'\cos\varphi}\right) \,.
 \label{chim}
 \end{eqnarray}
Here $v_0$ in Eq. (\ref{actionm}) is related to $\omega_0$ via the equation $\omega_0=
{v_0}/{r}+{\partial v_0}/{\partial r}$. Our goal now is to derive an effective action for
the zero harmonic, ${\cal I}_0+\Delta{\cal I}$ by integrating over all the other
harmonics,
 \begin{eqnarray}
 \exp\left(i{\cal I}_0+i\Delta{\cal I}\right)
 =\int \prod_{m>0}{\cal D}\omega_{\pm m}\, {\cal D}p_{\pm m}\,
 \exp\left(i{\cal I}\right)\,.
 \label{defin}
 \end{eqnarray}
The integration is Gaussian if to neglect third order terms in $p_m$ and $\omega_m$, as
explained above. Then $\Delta{\cal I}=\sum\Delta{\cal I}_m$ where
 \begin{eqnarray}
 e^{i\Delta{\cal I}_m} =\int{\cal D}\omega_{\pm m}\,
 {\cal D}p_{\pm m}\,e^{i{\cal I}_m+i{\cal I}_{-m}
 +i{\cal I}_{mm} +i{\cal I}_{im}}\,.
 \label{corrm}  \end{eqnarray}
If ${\cal I}_{im}=0$ then the expression (\ref{corrm}) is the normalization integral that
is equal to unity due to causality \cite{DP}. Therefore one can write
 \begin{eqnarray}
 \Delta{\cal I}=\sum\limits_{m>0}
 \sum\limits_{n=1}^\infty
 \frac{i^{n-1}}{n!} \left\langle
 \left({\cal I}_{im}\right)^n\right\rangle_c \,,
 \label{cormm}
 \end{eqnarray}
where the angular brackets mean integration over $\omega_{\pm m}$ and $p_{\pm m}$ with
the weight $\exp(i{\cal I}_m+i{\cal I}_{-m} +i{\cal I}_{mm})$ and the subscript $c$ means
an irreducible average (represented by connected diagrams).

Consistently considering small fluctuations (as in neglecting ${\cal I}_3$) we take only
the term with $n=1$ in Eq. (\ref{cormm}). We shall justify it later by observing that the
angular part remains small during the build-up of the strong fluctuation that we
consider. Therefore, the main object, that contributes to the $n=1$ term and need to be
examined, is the pair correlation function
 \begin{eqnarray}
 \left\langle \omega_m(t,r)
 \omega_{-m}(t',r')\right\rangle=F_m(t,t';r,r').
 \label{pair}
 \end{eqnarray}
The simultaneous pair correlation function satisfies the equation
 \begin{eqnarray}
 \partial_t F_m(t,t,r_1,r_2)
 +im\left[\frac{v_0(r_1)}{r_1} - \frac{v_0(r_2)}{r_2}\right]F_m(r_1,r_2)
 \nonumber \\
 -\partial_r\omega_0(r_1)\frac{i}{2}\int d r\,r^2
 \frac{u_1^{|m+1|}-u_1^{|m-1|}}{|r^2-r_1^2|}F_m(r,r_2)
 \label{EQPAIR} \\
 +\partial_r\omega_0(r_2)\frac{i}{2}\int d r\,r^2
 \frac{u_2^{|m+1|}-u_2^{|m-1|}}{|r^2-r_2^2|}F_m(r_1,r)
 =\chi_m(r_1,r_2) \,,
 \nonumber
 \end{eqnarray}
where $u_{1,2}=\mathrm{min} \{ {r}/{r_{1,2}}, {r_{1,2}}/{r}\}$. One should treat
separately the first angular harmonic, with $m=\pm1$.

 \subsection{Logarithmic approximation}

Let us pass to the logarithmic variable $\xi=\ln(r/L)$, where $L$ is the pumping length.
We are interested in small scales, $r\ll L$ where $|\xi|\gg1$. We consider only the
leading contributions in terms of large $|\xi|$. In this case, only the terms with $m=2$
are relevant in Eq. (\ref{cormm}) since the integration in the expressions
(\ref{Vel},\ref{EQPAIR}) is logarithmic only for them, as has been noticed already in
\cite{FL}, the feature is likely related to peculiarity of elliptic vortices in straining
flows \cite{Saffman}. Other harmonics behave as $r^{m-2}$ for $m>2$ i.e. are suppressed
exponentially in terms of the large logarithm $\xi$. In the logarithmic variables,
$\chi_m(\xi_1,\xi_2)$ for $m\neq0$ are nonzero only if both $|\xi_1|,|\xi_2| \lesssim1$,
since the integral (\ref{chim}) is zero for $r_1+r_2<L$ and decays as $L/\sqrt{r_1r_2}=
\exp[-(\xi_1+\xi_2)/2]$ for $r_1,r_2\to\infty$. That means that one can approximate
$\chi_m(\xi_1,\xi_2)\approx H_m \delta(\xi_1) \delta(\xi_2)$ for $m\not=0$. The zeroth
harmonics can be taken as $\chi_0(\xi_1,\xi_2) = H \theta(-\xi_1) \theta(-\xi_2)$ where
$\theta$ is the step function.

Next, we pass to the field $q(\xi)=r^2 p_0 (r)$. Then the bare action (\ref{actionzero})
for the zeroth harmonics is rewritten as
 \begin{eqnarray}
 {\cal I}_0= \int d t\, d \xi\, q \,
 \partial_t \, \omega_0
 +\frac{i}{2}\int d t\, d \xi_1\, d \xi_2\,
 \chi_0(\xi_1,\xi_2) q(\xi_1) q(\xi_2) \,.
 \label{semi1}
 \end{eqnarray}
The correction $\Delta{\cal I}$ (\ref{cormm}) in the main approximation (taking into
account only the term with $n=1$, $m=2$) can be written as
 \begin{eqnarray}
 \Delta{\cal I}\approx i \int d t\, d \xi\, q(t,\xi) \int_\xi d\zeta\,
 \left[F_2(t,t;\zeta,\xi)- F_2(t,t;\xi,\zeta)\right] \ .
 \label{semi2}
 \end{eqnarray}
Here the function $F_2$ has to be treated as a functional of $\omega_0$ to be extracted
from the equation (\ref{EQPAIR}) for $m=2$.

In terms of the logarithmic variable $\xi$ the equation (\ref{EQPAIR}) is rewritten as
 \begin{eqnarray}
 \partial_t F_2(\xi_1,\xi_2)
 +2i [w(\xi_1) - w(\xi_2)] F_2(\xi_1,\xi_2)-\chi_2(\xi_1,\xi_2)=
 \nonumber \\
 \frac{i\partial_\xi\omega_0(\xi_2)}{2}
 \left[\int_{\xi_2}^\infty\!\!\!\! d\xi\,
 F_2(\xi_1,\xi) +\int^{\xi_2}_{-\infty}\!\!\!\! d\xi\,
 e^{4(\xi-\xi_2)}F_2(\xi_1,\xi)\right]
 \label{eqpair2} \\
 -\frac{i\partial_\xi\omega_0(\xi_1)}{2} \left[
 \int_{\xi_1}^\infty\!\!\!\! d\xi\, F_2(\xi,\xi_2)
 +\int^{\xi_1}_{-\infty}\!\!\!\! d\xi\,
 e^{4(\xi-\xi_1)}F_2(\xi,\xi_2)\right] ,
 \nonumber
 \end{eqnarray}
where we denoted $w(\xi)\equiv {v_0}/{r}= \int^\xi_{-\infty} d\xi'\, \exp[2(\xi'-\xi)]
\omega_0(\xi')$. In the main logarithmic approximation we get from Eq. (\ref{eqpair2})
 \begin{eqnarray}
 \partial_t F_2(\xi_1,\xi_2)
 +i\omega_0(t,\xi_1)F_2(\xi_1,\xi_2)
 +\frac{i}{2}\partial_\xi\omega_0(t,\xi_1)
 \int_{\xi_1}^\infty d \xi\, F_2(\xi,\xi_2)
 \nonumber \\
 -i\omega_0(t,\xi_2)F_2(\xi_1,\xi_2)
 -\frac{i}{2}\partial_\xi\omega_0(t,\xi_2)
 \int_{\xi_2}^\infty d \xi\, F_2(\xi_1,\xi)
 =\chi_2(\xi_1,\xi_2) \,.
 \label{semi3}
 \end{eqnarray}
Putting in Eq. (\ref{semi3}) $\xi_1=\xi_2=\xi$ and substituting the result into the
expression (\ref{semi2}) one obtains
 \begin{equation}
 \Delta{\cal I}\approx -2 \int d t\, d \xi\, \frac{q}{\partial_\xi\omega_0}
 \partial_t F_2(\xi,\xi),
 \label{correction}
 \end{equation}
the term with $\chi_2$ is neglected since it is small in the interval of the direct
cascade.

 \subsection{Eigen functions}

The equation (\ref{semi3}) can be rewritten in the form
 \begin{eqnarray}
 \partial_t F_2(\xi_1,\xi_2)+
 i\hat O_1 F_2 -i\hat O_2 F_2=
 \chi_2(\xi_1,\xi_2) \,,
 \label{kernel1}
 \end{eqnarray}
where
 \begin{eqnarray}
 \hat O f(\xi)=\int d\zeta\,
 \left[\omega_0\delta(\xi-\zeta)
 +\frac{1}{2}\partial_\xi\omega_0\,\theta(\zeta-\xi)
 \right]f(\zeta) \,.
 \label{kernel2}
 \end{eqnarray}
Let us introduce eigen functions of the operator $\hat O$
 \begin{eqnarray}
 \hat O \varphi_\lambda=
  \omega_0(\xi) \varphi_\lambda(\xi) + \frac12
    \frac{\partial\omega_0(\xi)}{\partial\xi}
    \int\limits_{\xi}^\infty d\zeta \,
     \varphi_\lambda(\zeta)
   = \lambda \varphi_\lambda(\xi).
 \label{eigen}
 \end{eqnarray}
We assume (in accordance with the answer obtained) that $\omega_0(\xi)$ monotonically
diminishes from some value $s$ at $\xi=-\infty$ to zero at $\xi=+\infty$. Then a set of
the eigen functions of the operator $\hat O$ can be written as
 \begin{eqnarray}
 \varphi_\lambda=\theta(\omega_0-\lambda)
 2\partial_\xi\sqrt{\omega_0-\lambda}
 =\frac{\theta(\omega_0-\lambda)} {\sqrt{\omega_0-\lambda}}
 \partial_\xi \omega_0 ,
 \label{right}
 \end{eqnarray}
where $\theta$ is the step function and $0<\lambda<s$. The functions (\ref{right}) are
the right eigenfunctions of the operator $\hat O$. Analogously, one can define the left
eigenfunctions:
 \begin{eqnarray}
 \phi_\mu(\xi)
 = \frac{1}{2\pi} \lim_{\epsilon\to 0}
 {\rm Re}\left[\mu-\omega_0(\xi)
 +i\epsilon \right]^{-3/2},
 \label{left}
 \end{eqnarray}
where, again, $0<\mu<s$. The functions (\ref{left}) satisfy the equation
 \begin{eqnarray}
 \omega_0(\xi) \phi_\mu(\xi) + \frac{1}{2}
 \int^\xi_{-\infty} d\zeta\,
 \partial_\zeta\omega_0\phi_\mu(\zeta)=
 \mu \phi_\mu(\xi).
 \label{left1}
 \end{eqnarray}

The only thing which is important for what follows is the orthogonality and normality of
the right and left eigenfunctions. The factors in (\ref{right},\ref{left}) are chosen to
ensure the normalization condition that can be checked directly:
 \begin{eqnarray}
 \int d\xi\, \varphi_\lambda(\xi)\phi_\mu(\xi)
 =- \frac{1}{2\pi}\lim_{\epsilon\to 0} {\rm Re}
 \int_\lambda^s\!\! \frac{d\omega_0}{\sqrt{\omega_0-\lambda}}
 \frac{1}{(\mu-\omega_0+i\epsilon)^{3/2}}
 \nonumber \\
 =\frac{1}{\pi}\sqrt{s-\lambda}\lim_{\epsilon\to 0}
 {\rm Re}(\lambda-\mu-i\epsilon)^{-1}(\mu+i\epsilon
 -s)^{-1/2}=\delta(\mu-\lambda),
 \label{delta}
 \end{eqnarray}
provided $0<\mu,\lambda< s$. One can check completeness of the set
(\ref{right},\ref{left}):
 \begin{equation}
 \int_0^s d\lambda\
 \phi_\lambda(\xi) \varphi_\lambda(\zeta)
 =\delta(\xi-\zeta).
 \label{complete}
 \end{equation}
The check is reduced to an integral analogous to one (\ref{delta}).


Let us expand $F_2$ over the eigenfunctions (\ref{right}),
 \begin{eqnarray}
 F_2(t,\zeta_1,\zeta_2)= \int d\lambda_1\,d\lambda_2\,
 \Phi(t,\lambda_1,\lambda_2) \varphi_{\lambda_1}(\zeta_1)
 \varphi_{\lambda_2}(\zeta_2) ,
 \label{expand}
 \end{eqnarray}
where
 \begin{equation}
 \Phi(\lambda_1,\lambda_2)=
 \int d\xi_1\, d\xi_2\,
 \phi_{\lambda_1}(\xi_1)
 \phi_{\lambda_2}(\xi_2)
 F_2(\xi_1,\xi_2).
 \label{expand1}
 \end{equation}
Now we take into account that $\omega_0\to 0$ as $\xi\to+\infty$ and obtain:
 \begin{eqnarray}
 \left[\partial_t+i(\mu_1-\mu_2)\right] \Phi(t,\mu_1,\mu_2)
 +\int d\lambda_1\, \Phi(\lambda_1,\mu_2) J(\mu_1,\lambda_1)
 \nonumber \\
 +\int d\lambda_2\, \Phi(\mu_1,\lambda_2) J(\mu_2,\lambda_2)
 =\int d\xi_1\, d\xi_2\,
 \phi_{\mu_1}(\xi_1) \phi_{\mu_2}(\xi_2)
 \chi_2(\xi_1,\xi_2),
 \label{basic}
 \end{eqnarray}
where
 \begin{equation}
 J(\mu,\lambda)= \int d\zeta\, \phi_\mu(\zeta)
 \partial_t \varphi_\lambda(\zeta)
 = - \int d\zeta\, \partial_t \phi_\mu(\zeta)
 \varphi_\lambda(\zeta) .
 \label{con}
 \end{equation}
The equation (\ref{basic}) is equivalent to Eq. (\ref{kernel1}).

Substituting into the definition (\ref{con}) the explicit expressions
(\ref{right},\ref{left}) we get
 \begin{eqnarray}
 J(\mu,\lambda)= \frac{1}{\pi}\frac{\partial^2}{\partial\mu^2}
 \left\{\theta(\mu-\lambda)\int_\lambda^\mu \frac{d\omega_0\,\psi(t,\omega_0)}
 {\sqrt{(\omega_0-\lambda)(\mu-\omega_0)}}\right\} \,.
 \label{om1}
 \end{eqnarray}
Here we introduced the designation
 \begin{eqnarray}
 \partial_t\omega_0(t,\xi)=
 \psi(t,\omega_0) \,.
 \label{om2}
 \end{eqnarray}
Note that it follows from the definition (\ref{om2}) that
 \begin{eqnarray}
 \psi'=\frac{\partial\psi(t,\omega_0)}{\partial\omega_0}={\partial
 \psi\over\partial\xi}{\partial\xi \over\partial\omega_0}
 =\partial_t\ln(\partial_\xi\omega_0) \,.
 \label{om4} \end{eqnarray}
Performing the substitution $\omega_0=\lambda+(\mu-\lambda)x$ we get from Eq. (\ref{om1})
 \begin{eqnarray}
 J(\mu,\lambda)=\frac{\partial}{\partial\mu}
 \left[\delta(\mu-\lambda)\psi(\lambda)\right]
 +\frac{1}{2}\delta(\mu-\lambda)\psi'(\lambda)
 \nonumber \\
 +\frac{1}{\pi}\theta(\mu-\lambda) \int_0^1
 \frac{d x\,x^{3/2}}{\sqrt{1-x}}
 \psi''\left[\lambda+(\mu-\lambda)x\right]\,.
 \label{om3}
 \end{eqnarray}
We see that $J$ is the sum of singular terms and the term, which can be written as a
regular expansion over $\mu-\lambda$.

\subsection{Initial Condition}

As we will see, the instantonic solution for $\omega_0$ diminishes back in time.
Therefore the saddle-point approximation ceases to be correct at a time $t=t_*$ where
$\omega_0$ is of the order of typical (rms) fluctuation and one should pose the initial
condition for the instanton at $t=t_*$. We assume that the initial fluctuation (at
$t=t_*$) $\omega_0$ is some slow (logarithmic) function of the distances in the region
$|\xi|\lesssim \ln (L/r)$ so that $\partial_\xi\ln\omega_0\sim\xi^{-1}$. Next, we assume
that at $t=t_*$
 \begin{eqnarray}
 \langle\omega(\bm r_1)\omega(\bm r_2)\rangle
 \sim \left(H\gamma\right)^{2/3},
 \label{initial} \\
 \gamma=\ln\frac{|\bm r_1-\bm r_2|}{L}=\frac{1}{2}\ln\left[
 (r_1^2+r_2^2-2r_1 r_2\cos\varphi)/L^2\right]
 \nonumber \\
 =\frac{1}{2}\ln\left[\frac
 {r_1^2+r_2^2}{L^2}\right]
 +\frac{1}{2}\ln\left[1-
 \frac{2r_1 r_2\cos\varphi}{r_1^2+r_2^2}\right].
 \label{difference}
 \end{eqnarray}
It is important to stress that, strictly speaking, we cannot derive the second moment
(\ref{initial}) from the equation of motion. That choice is consistent with the flux
relation and, as we show below, is self-consistent with the higher moments described by
the PDF tail to be derived.

We now derive the second angular harmonic of the pair correlation function:
 \begin{eqnarray}
 F_2(t_*,\xi_1,\xi_2)=\int_0^{2\pi} \frac{d\varphi}{2\pi}
 \exp(-2i\varphi) \langle\omega(\bm r_1)\omega(\bm r_2)\rangle.
 \label{angular}
 \end{eqnarray}
If $|\xi_1|, |\xi_2| \gg1$ then a power-like function $f(\psi)$ can be expanded in the
ratio of the two last terms in Eq. (\ref{difference}). Then one obtains
 \begin{eqnarray}
 \int_0^{2\pi} \frac{d\varphi}{2\pi}
 \exp(-2i\varphi) f(\gamma)
 =-\frac{1}{4}\frac{d f}{d\gamma_0}
 \times
 \begin{array}{c}
 {r_2^2}/{r_1^2}\ \mathrm{if} \ r_1>r_2, \\
 {r_1^2}/{r_2^2}\ \mathrm{if} \ r_1<r_2,
 \end{array}
 \label{second}
 \end{eqnarray}
where $\gamma_0=(1/2)\ln[(r_1^2+r_2^2)/L^2]$ and we used the relation
 \begin{equation*}
 \int_0^{2\pi} dx\ \cos(2x)
 \ln(1-a\cos x)=
 \frac{\pi}{a^2}
 \left(a^2-2+2\sqrt{1-a^2}\right).
 \end{equation*}
We conclude that the function $F_2(t_*)$ exponentially in $\xi_1-\xi_2$ tends to zero as
$|\xi_1-\xi_2|\equiv |\ln(r_1/L)-\ln(r_2/L)|$ tends to infinity. Therefore it can be
estimated as $F_2(t_*)\sim H^{2/3} |\xi_1|^{-1/3}\delta(\xi_1-\xi_2)$.

Now we can analyze the initial value of the function $\Phi$
 \begin{equation}
 \Phi(t_*,\lambda_1,\lambda_2)=
 \int d\xi_1\, d\xi_2\,
 \phi_{\lambda_1}(\xi_1)
 \phi_{\lambda_2}(\xi_2)
 F_2(t_*,\xi_1,\xi_2).
 \label{expand2}
 \end{equation}
Substituting here the above estimate, one obtains
 \begin{equation}
 \Phi(t_*,\lambda_1,\lambda_2)\sim H^{2/3}
 \int_{-\infty}^0 d\xi\, |\xi|^{-1/3}
 \phi_{\lambda_1}(\xi)
 \phi_{\lambda_2}(\xi)
 . \label{initial1}
 \end{equation}
Again, the approximation (\ref{initial1}) implies $\lambda_1,\lambda_2\gg H^{1/3}$. What
is most important is that the integrand of (\ref{initial1}) has a third-order pole when
$\lambda_1\to\lambda_2\to \omega_0$. Integration over $\xi$ results in the second-order
pole: $\Phi_0(\lambda_1,\lambda_2)\propto (\lambda_1-\lambda_2)^{-2}$ at
$\lambda_1\to\lambda_2$. Therefore
 \begin{equation}
 \Phi(t_*,\lambda_1,\lambda_2) \sim
 \frac{(H\xi)^{2/3}}{\sqrt{\lambda_1\lambda_2}}
 \delta'(\lambda_1-\lambda_2),
 \label{deltaprime}
 \end{equation}
where $\xi$ is determined by the condition $\omega_0(t_*,\xi)=\lambda_1\approx\lambda_2$.

 \subsection{Adiabatic approximation}

The instantonic field $\omega_0(t,\xi)$ describes an optimal fluctuation that starts from
the rms level at the time $t_*$ and grows to a prescribed large value of $\varpi$. After
we find below the instanton solution $\omega_0(t,\xi)$, we see that its form changes slow
on its own rotation timescale $\omega_0^{-1}$. In particular, that means that the
eigenfunctions (\ref{right}) change slow too, so that one can neglect non-singular term
in the expression (\ref{om3}). Then
 \begin{eqnarray}
 J(\mu,\lambda)\to\frac{\partial}{\partial\mu}
 \left[\delta(\mu-\lambda)\psi(\lambda)\right]
 +\frac{1}{2}\delta(\mu-\lambda)\psi'(\lambda)\,.
 \label{om5}
 \end{eqnarray}
Substituting the expression (\ref{om5}) into Eq. (\ref{basic}) and omitting the
right-hand side we get
 \begin{eqnarray}
 \left[\partial_t+i(\mu_1-\mu_2)\right]\Phi(t,\mu_1,\mu_2)
 +\psi(\mu_1)\frac{\partial}{\partial\mu_1}\Phi
 +\psi(\mu_2)\frac{\partial}{\partial\mu_2}\Phi
 +\frac{3}{2}[\psi'(\mu_1)+\psi'(\mu_2)]\Phi=0 \,.
 \label{om6}
 \end{eqnarray}
The equation (\ref{om6}) can be solved by the method of characteristics. The equation for
the characteristics reads
 \begin{eqnarray}
 \frac{d\mu}{d t}=\psi(t,\mu)=
 \partial_t\omega_0(t,\xi) \,,
 \label{om7}
 \end{eqnarray}
where $\xi$ is a function of $\mu$ to be extracted from the condition $\mu=\omega_0$. An
obvious solution of the equation (\ref{om7}) is
 \begin{eqnarray}
 \mu(t)=\omega_0(t,\xi) \,,
 \label{om8} \end{eqnarray}
where $\xi$ plays the role of the marker of the characteristic. A solution of the
equation (\ref{om6}) is written as
 \begin{eqnarray}
 \Phi(t,\mu_1,\mu_2)= \Phi[t_*,\mu_1(t_*),\mu_2(t_*)]
 \nonumber \\
 \times \exp\left(\int_{t_*}^t d s\, \left\{-i\mu_1(s)+i\mu_2(s)
 -\frac{3}{2}\psi'[s,\mu_1(s)]
 -\frac{3}{2}\psi'[s,\mu_2(s)]\right\}\right) \,.
 \label{om9}
 \end{eqnarray}
Using Eq. (\ref{om4}) we get
 \begin{eqnarray}
 \Phi(t,\mu_1,\mu_2)=
 \left[\frac{\partial_\zeta\omega_0(t_*,\zeta_1)}
 {\partial_\zeta\omega_0(t,\zeta_1)}\,
 \frac{\partial_\zeta\omega_0(t_*,\zeta_2)}
 {\partial_\zeta\omega_0(t,\zeta_2)}\right]^{3/2}
 \Phi[t_*,\mu_1(t_*),\mu_2(t_*)]
 \nonumber \\ \times
 \exp\left\{\int_{t_*}^t d s\,
 \left[-i\mu_1(s)+i\mu_2(s)\right]\right\} \,,
 \label{om10}
 \end{eqnarray}
where the variables $\zeta_1$, $\zeta_2$ have to be extracted from the relations
$\mu_1=\omega_0(t,\zeta_1)$, $\mu_2=\omega_0(t,\zeta_2)$.

Now we can estimate a role of the regular contribution omitted in Eq. (\ref{om5}), see
Eq. (\ref{om3}). Substituting the expression (\ref{om10}) into Eq. (\ref{basic}) we then
conclude that an integration over $\lambda_1$ or over $\lambda_2$ in the omitted terms is
determined mainly by the oscillating factor in Eq. (\ref{om10}). Then $\int
d\lambda_1\to(t-t_*)^{-1}$ (and the same for the integration over $\lambda_2$) and
therefore the omitted terms in the equation (\ref{basic}) can be estimated as
 \begin{eqnarray}
 \frac{1}{t-t_*} \psi''(\mu_1) \Phi(\mu_1,\mu_2) \,.
 \nonumber \end{eqnarray}
Comparing the term with Eq. (\ref{om6}), we conclude, that the omitted terms have an
additional small factor $[(t-t_*)\omega_0]^{-1}$ and therefore the approximation leading
to Eq. (\ref{om6}) is correct.

Let us calculate $F_2(\xi_1,\xi_2)$ using the expression (\ref{om10}):
 \begin{eqnarray}
 F_2(t,\xi_1,\xi_2) = \int d \mu_1 d \mu_2
  \frac{\theta[\omega(t,\xi_1)-\mu_1]}{\sqrt{\omega(t,\xi_1)-\mu_1}}
  \frac{\theta[\omega(t,\xi_2)-\mu_2]}{\sqrt{\omega(t,\xi_2)-\mu_2}}
 \label{om11} \\ \times
 \partial_\xi\omega_0(t,\xi_1) \partial_\xi\omega_0(t,\xi_2)
 \left[\frac{\partial_\zeta\omega_0(t_*,\zeta_1)}
    {\partial_\zeta\omega_0(t,\zeta_1)}\,
    \frac{\partial_\zeta\omega_0(t_*,\zeta_2)}
    {\partial_\zeta\omega_0(t,\zeta_2)}\right]^{3/2}
    \nonumber \\ \times
 \Phi[t_*,\omega_0(t_*,\zeta_1),\omega_0(t_*,\zeta_2)]
    \exp\left\{\int_{t_*}^t d s\,
    \left[-i\omega_0(s,\zeta_1)
    +i\omega_0(s,\zeta_2)\right]\right\} \,.
 \nonumber
 \end{eqnarray}
Recall, that $\zeta_1$ and $\zeta_2$ have to be extracted  from the relations
$\mu_1=\omega_0(t,\zeta_1)$, $\mu_2=\omega_0(t,\zeta_2)$. After substituting
(\ref{deltaprime}) into (\ref{om11}), in the integral over $\mu_2$ we keep only the pole
term at $\mu_2\to\mu_1$, $\zeta_2\to\zeta_1$; differentiating the exponent gives the
$t-t_*$ factor because of the slowness of the instanton. Passing then from the
integration over $\mu_1$ to ones over $\zeta$ we obtain
 \begin{eqnarray}
 F_2(t,\xi,\xi) \sim
 H^{2/3}(t-t_*) [\partial_\xi\omega_0(t,\xi)]^2\int d \zeta \, \zeta^{2/3}
 \partial_\zeta \ln \omega_0(t_*,\zeta)
 \frac{\theta[\omega_0(t,\xi)-\omega_0(t,\zeta)]}{\omega_0(t,\xi)-\omega_0(t,\zeta)}
 \nonumber \\\sim(t-t_*)(\partial_\xi\omega_0)
 H^{2/3}\xi^{-1/3}\ln(\omega_0/\partial_\xi\omega_0) \ .
 \label{final0}
 \end{eqnarray}
Here we used $\partial_\zeta \ln \omega_0(t_*,\zeta)\simeq 1/\zeta$, it is equivalent to
assuming that the initial $\omega_0(\xi)$ is a power-like function of the logarithm
$\xi$. The logarithmic divergence in (\ref{final0}) is cut off due to a finite
(order-unity) width of $F_2(t_*,\xi_1,\xi_2)$ over $\xi_1-\xi_2$.

Now we turn to the pumping contribution:
\begin{eqnarray}
 \Phi(\mu,\lambda,t)=  \int_0^{t-t_*}d\tau\
 \exp\left[-i(\mu-\lambda)(t-t_*-\tau)\right]
 \Xi(\mu,\lambda),
 \label{adiab} \\
 \Xi(\mu,\lambda)=\int d\xi_1\, d\xi_2\,
 \phi_{\mu}(\xi_1) \phi_{\lambda}(\xi_2)
 \chi_2(\xi_1,\xi_2).
 \label{pumpingt}
 \end{eqnarray}
The function $\chi_2(\xi_1,\xi_2)$ is nonzero provided $\xi_1, \xi_2\sim 1$. Since the
integration over $\xi_1$ and $\xi_2$ in Eq. (\ref{pumpingt}) smears the singularities in
$\phi$ then $\Xi$ a smooth function of $\mu$ and $\lambda$. Next, a characteristic
$\omega_0$ in the integral (\ref{pumpingt}) is $H^{1/3}$. Therefore $\Xi\propto
\mu^{-3/2}$ at $\mu\gg H^{1/3}$ and $\Xi\propto \lambda^{-3/2}$ at $\lambda\gg H^{1/3}$.
Thus the characteristic $\mu$ and $\lambda$ in the integral (\ref{expand}) are less or of
the order of $H^{1/3}$. Then we obtain
 \begin{eqnarray}
 F_2^\mathrm{pump}(\xi,\xi)\approx
 \frac{(\partial_\xi\omega_0)^2}{\omega_0}
 \int d\mu\ d\lambda\ \int_0^{t-t_*}d\tau\
 \nonumber \\
 \exp\left[-i(\mu-\lambda)(t-t_*-\tau)\right]
 \Xi(\mu,\lambda)
 \sim H^{1/3}\frac{(\partial_\xi\omega_0)^2}{\omega_0}.
 \label{pumpingc}
 \end{eqnarray}
Since we shall obtain a slow instanton with $H^{1/3}t_*\gg1$, the contribution
(\ref{final0}) is larger than (\ref{pumpingc}). That means that the pumping-produced
anisotropic fluctuations give lesser contribution than deformation of an initial
fluctuation. The consequence is that the tail of the vorticity PDF is insensitive to the
form of the pumping correlation function and is determined solely by its zeroth moment
i.e. the vorticity flux. That means universality of the statistics of strong vorticity
fluctuations.

The estimation (\ref{final0}) determines the main contribution to $F_2(\xi_1,\xi_2)$
where $\xi_1\sim \xi_2$. We checked this time dependence of $F_2$ by solving numerically
Eq. (\ref{semi3}) using different time-independent $\omega_0$. At the beginning, we have
chosen $F_2$ being determined by a typical fluctuation. Then we checked that $F_2$
linearly grows as time runs. It is interesting to note that the antisymmetric in $\xi_1$,
$\xi_2$ part in $F_2$ saturates. The behavior is in accordance with Eq. (\ref{semi3}).

 \section{Instanton}
 \label{sec:inst}

Substituting the expression (\ref{final0}) into Eq. (\ref{correction}) one obtains
 \begin{eqnarray}
 \Delta{\cal I} \sim   -H^{2/3}\int d t\, d\xi\, q
 \xi^{-1/3} (\partial_\xi\omega_0)^{-1}\partial_t
 [(t-t_*) \partial_\xi\omega_0\ln(\omega_0/\partial_\xi\omega_0)]  \,.
 \label{self4}
 \end{eqnarray}
Collecting the expressions (\ref{semi1},\ref{self4}) we get finally the effective action
 \begin{eqnarray}
 {\cal I}_{\rm eff}= \int d t\, d \xi\, q \, \left\{\partial_t \,
 \omega_0- c H^{2/3}\xi^{-1/3} (\partial_\xi\omega_0)^{-1} \partial_t
 [(t-t_*) \partial_\xi\omega_0\ln(\omega_0/\partial_\xi\omega_0)] \right\}
 +\frac{i}{2}H\int d t\, d \xi_1\, d \xi_2\,
 q(\xi_1) q(\xi_2) \,,
 \nonumber
 \end{eqnarray}
where $c\sim1$ and we substituted $\chi_0(\xi_1,\xi_2) = H \theta(-\xi_1)
\theta(-\xi_2)$. By rescaling $q$, $\omega_0$ and $H$ we can put $c\to1$.
 \begin{eqnarray}
 {\cal I}_{\rm eff}= \int d t\, d \xi\, q \, \left\{\partial_t \,
 \omega_0- H^{2/3}\xi^{-1/3}(\partial_\xi\omega_0)^{-1}\partial_t \left[
 (t-t_*) \partial_\xi\omega_0\ln(\omega_0/\partial_\xi\omega_0)\right] \right\}
 \nonumber \\
 +\frac{i}{2}H\int d t\, d \xi_1\, d \xi_2\,
 q(\xi_1) q(\xi_2) \,,
 \label{apo1}
 \end{eqnarray}

We are interested in the PDF of $\varpi$, that is $\omega$ coarse-grained over the scale
$R$. In our terms, it can be written as
 \begin{equation}
 \varpi=\frac{2}{R^2}\int_0^R dr\, r\, \omega_0
 \approx \omega_0 [\ln(R/L)],
 \label{coarse}
 \end{equation}
because of the logarithmic character of the $\omega_0$ dependence on $r$. Thus, we should
fix $\omega_0 [\ln(R/L)]=\varpi$ at the observation time. Since we consider steady-state
turbulence, the moment of measurement is arbitrary, we choose it to be $t=0$. Then PDF
${\cal P}(\varpi)$ can be calculated as the path integral
 \begin{equation}
 {\cal P}=\int {\cal D}\omega_0\ {\cal D}q \
 \exp\left(i{\cal I}_{\rm eff}\right),
 \label{pdfint}
 \end{equation}
taken at the condition $\omega_0[0,\ln(r/L)]=\varpi$ and for the fields $\omega_0$, $q$
defined at negative times $t$ \cite{FKLM}. The last property is explained by causality:
the values of the fields at positive times cannot influence the PDF ${\cal P}(\varpi)$.
Since $\omega_0$ is fixed at a single point at $t=0$ the field $q$ at $t=0$ satisfies
$q(\zeta)\propto \delta[\xi-\ln(R/L)]$ that reflects the measuring procedure.

In the saddle-point approximation, we put $\ln{\cal P}\approx i{\cal I}_{\rm
eff}^\mathrm{extr}$ where ${\cal I}_{\rm eff}^\mathrm{extr}$ is the extremum value of the
effective action. The extremum conditions $\delta{\cal I}_{\rm eff}/\delta
\omega_0=0=\delta{\cal I}_{\rm eff}/\delta q$ give the so-called instanton equations:
 \begin{eqnarray}
 \partial_t\omega_0=  H^{2/3}
 \zeta^{-1/3}(\partial_\zeta\omega_{0})^{-1}
 \partial_t\left[(t-t_*) \partial_\zeta
 \omega_{0}\ln(\omega_0/\partial_\zeta\omega_{0})\right]+ H Q(t) \,,
 \label{inst4} \\
 \partial_tq+H^{2/3}\partial_\zeta \left\{\ln(\omega_0/\partial_\zeta\omega_0)
 \zeta^{-1/3}(\partial_\zeta\omega_{0})^{-1} \partial_t [(t-t_*) q]\right\}=0\,,
 \label{inst4q}
 \end{eqnarray}
where $Q(t)=-i\int d\zeta\,q(\zeta,t)$. In deriving Eq. (\ref{inst4q}) we exploited large
value of the logarithm $\ln(\omega_0/\partial_\zeta\omega_0)\simeq\ln|\zeta|$ so that in
the main order we only account for the terms in the equations that contain the logarithm.
Apart from the logarithm, the correction (\ref{self4}) depends only on the vorticity
spatial derivative $\partial_\xi \omega_{0}$. As a result, the variation with respect to
vorticity gives the equation (\ref{inst4q}), which has the form of a continuity equation,
so that $dQ/dt=0$ in the main order. We see from Eq. (\ref{inst4}) that the first term in
the right-hand side is negative at $\zeta<0$ that is the correction (\ref{self4})
describes decrease of the vorticity due to deformation of the circular vortex by elliptic
perturbations. Substituting the relation (\ref{inst4}) into the expression (\ref{apo1}),
one finds
 \begin{equation}
 \ln {\cal P} \approx -\frac{H}{2}
 \int dt \ Q^2 \approx -\frac{H}{2} Q^2 |t_*|.
 \label{lnp}
 \end{equation}

Since $Q$ is $t$-independent, one readily obtains from Eq. (\ref{inst4}) that $\omega_0$
grows linearly with time, $\omega_0(t,\xi)=\beta(\xi)\cdot(t-t_*)$. Then we find for the
factor $\beta$
 \begin{equation}
 \beta=2 \frac{H^{2/3}}{\xi^{1/3}}
 \ln\frac{\beta}{\partial_\xi\beta} +HQ.
 \label{beta1}
 \end{equation}
Replacing here the ratio $\beta/\partial_\xi\beta$ by $\xi$ one obtains
 \begin{eqnarray}
 \omega_0=\left[2 \frac{H^{2/3}}{\xi^{1/3}}
 \ln|\xi| +HQ\right] (t-t_*),
 \label{beta2} \\
 \varpi=\left[-2\frac{H^{2/3}}{[\ln(L/R)]^{1/3}}
 \ln\ln(L/r) +HQ\right] |t_*|.
 \label{beta3}
 \end{eqnarray}
Then one obtains from Eq. (\ref{lnp})
 \begin{equation}
 \ln {\cal P} \simeq -
 \frac{Q^2 \varpi}{2\{Q-2H^{-1/3}[\ln(L/R)]^{-1/3}
 \ln\ln(L/R)\}}.
 \label{beta4}
 \end{equation}
Optimizing the expression over $Q$ one gets
 \begin{eqnarray}
 \ln {\cal P} \simeq -4 H^{-1/3}[\ln(L/R)]^{-1/3}
 \ln\ln(L/r) \varpi
 \label{beta5} \\
 \varpi=2\frac{H^{2/3}}{[\ln(L/R)]^{1/3}}
 \ln\ln(L/r) |t_*|.
 \label{beta6}
 \end{eqnarray}
The  value of $\omega_0(t_*)$ does not matter with logarithmic accuracy as long it is
much smaller than $\varpi$. The expression (\ref{beta5}) leads to the final answer
(\ref{answer0}) where we omitted the slow factor $\ln[\ln(L/R)]$.

We can use the instanton solution found to check the validity of all the assumptions made
in the derivation of the effective action. Remind that we consider the case $\ln(L/R)
\gg1$. The applicability condition of the saddle-point approximation is
 \begin{equation}
 |\varpi^3|\gg H \ln(L/R) \,.
 \label{es1}
 \end{equation}
The fluctuations on the background of our instanton are indeed small: using the instanton
solution we estimate $F_2 \simeq t_*\omega_0\xi^{-4/3}\omega_0^2/\xi\ll\omega_0^2$ as was
assumed. That justifies neglecting ${\cal L}_3$ and $n>1$ terms in (\ref{cormm}). Let us
now estimate $F_m$ for $m>2$ and compare it with $F_2$. Zero mode of (\ref{EQPAIR}) must
allow cancelation of integral and non-integral terms which is possible only when the
integrals in (\ref{EQPAIR}) are logarithmic. That requires $F_m\propto r^{m-2}\propto
\exp[(2-m)\xi]$, therefore those terms are exponentially suppressed comparing to $F_2$.
The instanton duration time is such that $|\varpi t_*|\gg1$ so that our instanton is
``slow'', that is indeed $\omega_0(t)$ changes slowly comparing to itself.

 \section{Discussion}
 \label{sec:discussion}

It is  illuminating to compare vorticity statistics in the direct $2d$ cascade with the
statistics of the passive scalar in a spatially smooth random flow
\cite{CFG,Kra67,Leith,Batch}. For a passive scalar $\theta$ coarse-grained over a scale
$R$ less than the pumping length $L$ one can get the asymptotic behavior of the
single-point PDF in a smooth random flow by the following simple reasoning. Large values
of $\theta$ are achieved when there is no stretching for a time which is much longer than
the mean stretching time $\lambda^{-1}\ln(L/R)$, where $\lambda$ is the Lyapunov
exponent. During that time, the passive scalar is pumped by a random forcing, i.e. it has
Gaussian statistics with the linearly growing variance:
 \begin{equation}
 {\cal P}(\theta)\sim \int dt\ {\cal Q}(t)\exp(-\theta^2/Pt)\
 ,\label{theta1}
 \end{equation}
where ${\cal Q}(t)$ is the probability of no stretching during time $t$. Stretching is
correlated on the velocity timescale $\lambda_0^{-1}$, which is independent of $\theta$.
For every stretching event, the scalar blob is stretched by order $e$ and we ask for the
probability that there were less than the number $\ln Pe$ such events during $t$. For
$t\gg\lambda_0^{-1}$, this is the probability of the Poisson process $\ln{\cal Q}(t)\sim
-\lambda_0 t +O[\ln(L/R)]$. Doing a saddle-point integration over $t$ we obtain the
exponential tail (first suggested in \cite{SS1,CFKLa} and derived by the instanton
formalism in \cite{FKLM,BGK}):
 \begin{equation}
 \ln[{\cal P}(\theta)]\sim
 -\theta\sqrt{\lambda_0/P}+O[\ln(L/R)].
 \label{theta2}
 \end{equation}

In the work \cite{FL}, we established some features of the direct vorticity cascade using
an analogy between the vorticity $\omega$ and passive scalar $\theta$. Developing that
analogy further, one can propose an interpretation of the tail (\ref{answer0}). For the
vorticity cascade, we may use similar reasoning as for the passive scalar with the
knowledge added from \cite{FL} that the stretching correlation time is the mean total
stretching time $\ln^{2/3}(L/R)/H^{1/3}$ from the scale $R$ to $L$. That gives
 \begin{equation}
 {\cal P}(\varpi)\sim \int dt\
 {\cal Q}(t,\varpi)\exp(-\varpi^2/Ht)\sim \int dt\
 \exp\left[-\varpi^2/Ht -tH^{1/3}\ln^{-2/3}(L/R)\right] \ .
 \label{omega1}
 \end{equation}
The saddle-point integration shows that the main contribution comes from $t\sim
\varpi\ln^{1/3}(L/R)H^{-2/3}$ in agreement with (\ref{beta6}), and the result $\ln[{\cal
P}(\varpi)]\simeq - |\omega|/|H\ln(L/R)|^{1/3}$  reproduces the dependence of
(\ref{answer0}). We see that vorticity is indeed like passive scalar: the stronger the
fluctuation the longer it lives which leads to a sub-Gaussian PDF tail. Such tails were
observed in numerical simulations \cite{Claudia,Bof}. For quantities (like velocity)
whose statistics is determined by fast events, their PDF have tails steeper than Gaussian
\cite{97FL}.

In a finite box, coherent vortices may appear due to an inverse cascade
\cite{07CCKL,08XPFS,09XSF}. The vortices have a well-defined spatial profile of the
average velocity field that is explained by an interplay of the average profile time
derivative (or friction) and an effective pumping related to long-correlated fluctuations
\cite{09CKL}. An interesting question that is a subject of future investigations concerns
an influence of the coherent vortices on the enstrophy cascade. One can also think about
extension of our instantonic approach to the inverse (energy) cascade of the $2d$
turbulence.

\acknowledgements

We thank I. Kololokolov and S. Korshunov for valuable discussions. The work was supported
by grants of Minerva and Israeli Science Foundations, by the grant no. 09-02-01346-a,
state contract 02.740.11.5195 and FTP ``Kadry" by MES of Russian Federation.

\end{document}